\documentclass[]{sciarticle}
\usepackage{graphicx}
\usepackage{multirow}

\renewcommand{\vec}{\mathbf}
\renewcommand{\vec}{}

\begin{document}

\title{
Helium atom excitations by the $GW$ and Bethe-Salpeter many-body formalism
}

\author{Jing Li}
\affiliation{Universit{\'e} Grenoble Alpes, 38000 Grenoble, France}
\affiliation{CNRS, Institut N\'eel, 38042 Grenoble, France}
\author{Markus Holzmann}
\affiliation{Universit{\'e} Grenoble Alpes, 38000 Grenoble, France}
\affiliation{CNRS, LPMMC, 38042 Grenoble, France}
\affiliation{European Theoretical Spectroscopy Facility (ETSF)}
\author{Ivan Duchemin}
\affiliation{Universit{\'e} Grenoble Alpes, 38000 Grenoble, France}
\affiliation{CEA, INAC-MEM L\_Sim, 38054 Grenoble, France}
\author{Xavier Blase}
\affiliation{Universit{\'e} Grenoble Alpes, 38000 Grenoble, France}
\affiliation{CNRS, Institut N\'eel, 38042 Grenoble, France}
\author{Valerio Olevano}
\affiliation{Universit{\'e} Grenoble Alpes, 38000 Grenoble, France}
\affiliation{CNRS, Institut N\'eel, 38042 Grenoble, France}
\affiliation{European Theoretical Spectroscopy Facility (ETSF)}

\date{\today}

\begin{abstract}
Helium atom is the simplest many-body electronic system provided by nature.
The exact solution to the Schr\"odinger equation is known for helium ground and excited states, and represents a workbench for any many-body methodology.
Here, we check the \textit{ab initio} many-body $GW$ approximation and Bethe-Salpeter equation (BSE) against the exact solution for helium.
Starting from Hartree-Fock, we show that $GW$ and BSE yield impressively accurate results on excitation energies and oscillator strength, systematically improving time-dependent Hartree-Fock.
These findings suggest that the accuracy of BSE and $GW$ approximations is not significantly limited by self-interaction and self-screening problems even in this few electron limit.
We further discuss our results in comparison to those obtained by time-dependent density-functional theory.
\end{abstract}

\maketitle

\paragraph{\textbf{Introduction - }}
The solution of the Schr\"odinger equation in interacting many-body  systems is a formidable problem in condensed matter and nuclear physics, as well as in quantum chemistry. 
Starting with the Hartree-Fock (HF) method as one of the first approaches to treat the quantum many-body problem, different theoretical formalisms have been developed over time \cite{MRCbook}:
approaches relying on the many-body wave function, like quantum Monte Carlo or quantum chemistry methods, or on the electronic density, like density-functional theory (DFT) in its static or time-dependent (TDDFT) form or,
finally, quantum field theoretical, Green function based methods.
While exact in principle, all of these methods rely in practical calculations on approximations and recipes whose validity are difficult to judge.

The $GW$ approximation \cite{Hedin65,StrinatiHanke80,StrinatiHanke82,HybertsenLouie85,GodbySham87} and Bethe-Salpeter equation (BSE) \cite{SalpeterBethe51,HankeSham79,OnidaAndreoni95} are \textit{ab initio} approaches to calculate the electronic structure within many-body Green function theory.
The $GW$ approximation of the self-energy was first applied to the homogeneous electron gas or jellium model \cite{Hedin65} and later extended to real solids \cite{StrinatiHanke80,HybertsenLouie85,GodbySham87} and molecules \cite{BlaseOlevano11,Bruneval12,VanSettenRinke15}.
The Bethe-Salpeter equation was originally developed to describe nuclear two-particle bound states \cite{SalpeterBethe51} and later applied to calculate optical spectra in solids \cite{HankeSham79,AlbrechtOnida98,RohlfingLouie98,BenedictBohn98} and excitations in molecules \cite{Jac15a,BrunevalNeaton15,BaumeierRohlfing12}.
Judged by comparison with experiment \cite{hedin99}, $GW$ and BSE provide reasonably accurate results for electronic excitations. 
However, benchmarks against the experiment, as done so far for $GW$ and BSE, are always affected by unaccounted effects not present in the theoretical description (non-Born-Oppenheimer, electron-phonon, relativistic corrections, etc.), which can mask the real many-body performances of the approaches.
Validation of the approximations against exact analytic solutions---or at least accurate, safe, and reliable numerical solutions---in realistic models or simple real systems is an unavoidable step for further improvement.

Neutral helium atom is the simplest real many-electron system in nature, at the extreme limit where ``many electron'' is reduced to only two electrons.
It is the lightest atom presenting the complication of the many-body correlations, and, due to the full rotational symmetry of the ionic potential, it is simpler than the hydrogen molecule. 
This allows us to write the electronic wave functions on a basis of only three scalar coordinates, Hylleraas functions, whose parameters are then varied to obtain the ground or the excited states.
Calculations can achieve extremely high accuracy, $10^{-7}$ Ha on energies in the best cases \cite{KonoHattori84}, and their quantitative agreement with the experiment may be considered as one of the triumphs of quantum mechanics.
The exact theoretical result makes helium, that is, almost a toy model and yet a real system, an ideal workbench for any many-body methodology.

The purpose of this work is to check the validity of many-body $GW$ and BSE calculations of the helium excitation spectrum against the exact results of the idealized nonrelativistic helium Hamiltonian \cite{KonoHattori84}.
This is a check of the core of the methodology, the solution to the many-body problem.
Helium atom can represent a very severe test case for these many-body theories whose underlying approximations have been devised to describe the extreme opposite limit of many electrons, like bulk solids, where screening is much more important.
The concept itself of screening is the way towards correlations beyond Hartree-Fock by the $GW$ approximation.
The validity of the BSE two-particle electron-hole propagator might be questioned in a system where the hole is dug in a Fermi sea of only two electrons.
Important drawbacks of the $GW$ and BSE approaches, like the self-interaction \cite{NelsonGodby07,MorrisGodby07,Fernandez09,ChangJin12} or the self-screening problem \cite{RomanielloReining09,AryasetiawanKarlsson12}, should directly manifest in helium atom calculations and strongly limit the accuracy of the results.
On the other hand, the validation of $GW$ and BSE on helium in atomic physics represents an important confirmation of this methodology;
it quantifies performances and errors in condensed matter;
it may face reluctance and skepticism within theoretical chemistry where $GW$ and BSE were recently exported;
it may receive attention from nuclear physics for application to excitations in nuclei.

The high accuracy of experimental spectroscopy on helium  have pushed theoretical calculations of higher-order effects, like nucleus finite mass recoil, relativistic fine structure, and QED radiative corrections \cite{PachuckiSapirstein03,PachuckiYerokhin10,ZhangDrake96}.
Furthermore, the electron-hole (\textit{e-h}) Bethe-Salpeter calculation of the present paper should not be confused with the electron-electron extended, three-body external potential, relativistic Bethe-Salpeter calculations \cite{ZhangDrake96,Zhang96,Zhang97} of higher-order corrections to the fine structure.
Although an active research field, these corrections are not the purpose of the present work.
Unless otherwise stated, we use atomic units in the following.

\begin{figure}
 \includegraphics[width=\columnwidth]{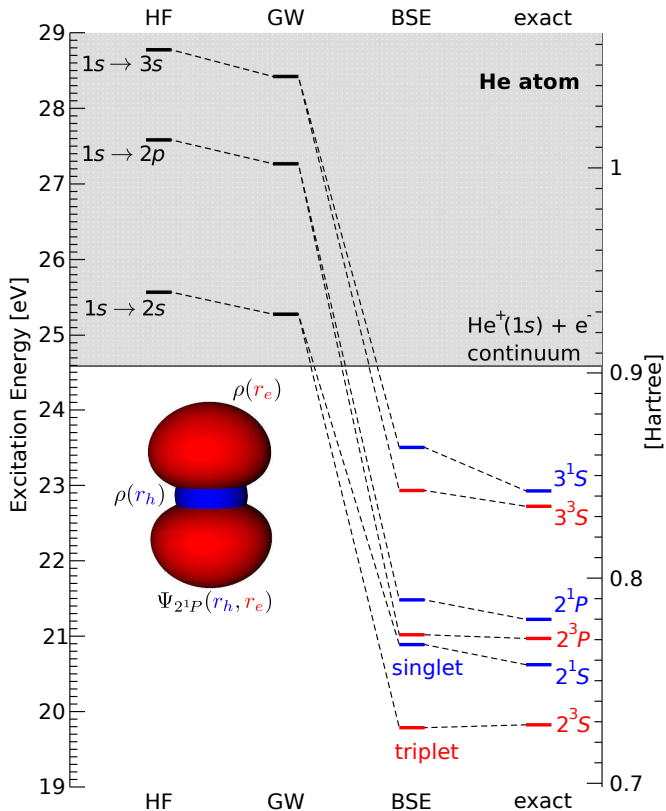}
 \caption{
   He atom excitation spectrum.
   The 0 of the energy is set to the ground state $1^1\!S$. 
   Inset: $2^1\!P$ electron-averaged hole (blue) and hole-averaged electron (red) distributions, isosurfaces taken at $5 \times 10^{-4}$ a.u.
 }
 \label{helevels}
\end{figure}

\paragraph{\textbf{Calculation - }}
The many-body perturbation theory methodology is a three step procedure.
(1) First, the calculation of a starting, approximated electronic structure (both energies and wave functions) of a somewhat arbitrary Hamiltonian chosen as  zero order.
Today, the most common choice is DFT, e.g., in the local-density approximation (LDA) or beyond.
Here, we preferred, rather, a return to the origins \cite{StrinatiHanke80,StrinatiHanke82,HankeSham79} and chose Hartree-Fock.
This choice avoids hybridization with alternative theories, like DFT, so to allow, at the end, a clear comparison between $GW$-BSE and DFT-TDDFT results.
In fact, the accuracy of DFT-TDDFT results on He critically relies on the first DFT step, and choice of the same initial step here too can raise doubts about the effective merit of the following $GW$ and BSE steps.
We then calculate the HF electronic structure energies and wave functions, $\epsilon^\mathrm{HF}_i, \phi^\mathrm{HF}_i(\vec{r})$,
\[
  H_\mathrm{H}(\vec{r}) \phi^\mathrm{HF}_i(\vec{r}) + \int d\vec{r}' \, 
     \Sigma_x(\vec{r},\vec{r}') \phi^\mathrm{HF}_i(\vec{r}')
  = \epsilon^\mathrm{HF}_i \phi^\mathrm{HF}_i(\vec{r})
  ,
\]
where $H_\mathrm{H}(\vec{r}) = - \partial_\vec{r}^2/2 + v_\mathrm{ext}(r) + v_\mathrm{H}(\vec{r})$ is the Hartree Hamiltonian and $\Sigma_x$ is the Fock exchange operator.
We use the full nuclear potential $v_\mathrm{ext}(r) = - Z / r$ and perform an all-electron calculation to reduce sources of inaccuracy  related to pseudopotential issues in our comparison to the exact result.
(2) The starting electronic structure is then used to calculate the Green function $G$, the dynamically screened interaction, $W$, in the random-phase approximation, and the self-energy, $\Sigma$, that, in the $GW$ approximation, is the convolution product of $G$ and $W$:
\begin{equation}
  \Sigma(\vec{r},\vec{r}',\omega) = \frac{i}{2\pi} \int d\omega' \, G(\vec{r},\vec{r}',\omega-\omega') W(\vec{r},\vec{r}',\omega')
  .
  \label{sigmagw}
\end{equation}
The $GW$ electronic structure, $\varepsilon_i, \varphi_i(\vec{r})$, is then calculated by solving the quasiparticle equation,
\begin{equation}
  H_\mathrm{H}(\vec{r}) \varphi_i(\vec{r}) + \int d\vec{r}' \, 
      \Sigma(\vec{r},\vec{r}',\omega=\varepsilon_i) \varphi_i(\vec{r}')
  = \varepsilon_i \varphi_i(\vec{r}).
\label{qpeq}
\end{equation}
In self-consistent (SC) $GW$, the new electronic structure is used to iterate the calculation of $G$, $W$, and $\Sigma$ Eq.~(\ref{sigmagw}) and solve Eq.~(\ref{qpeq}), until convergence is achieved.
There are several ways to perform full or ``quenched'' SC-$GW$ \cite{HolmVonBarth98,FaleevKotani04,BrunevalReining06,StanVanLeeuwen09}.
The question of which is better and even whether the SC form really improves on the spectral properties \cite{HolmVonBarth98} is still open.
Here, we want to check the standard procedure, so we refer to one iteration, non-self-consistent $G_0W_0$, or apply self-consistency only on the energies $\varepsilon_i$, keeping unchanged the wave functions $\varphi_i(\vec{r})$. 
Indeed, the difference between HF and exact wave functions has been found to be barely discernible for He \cite{UmrigarGonze94}.
(3) The final step is the resolution of the Bethe-Salpeter equation on top of the $GW$ electronic structure.
This is equivalently done by solving for the eigenvalues $E_\lambda$ and the eigenvectors $\Psi_\lambda$ of an excitonic Hamiltonian,
\begin{equation}
 \left(
 \begin{array}{cc}
   H^R & H^C \\
   - H^{C^*} & -H^{R^*} \\
 \end{array}
 \right)
 \Psi_\lambda = E_\lambda \Psi_\lambda
 , \label{bseeq}
\end{equation}
defining the resonant ($R$) and coupling ($C$) parts as
\begin{eqnarray*}
  H^R_{vc,v'c'} &=& (\varepsilon_c - \varepsilon_v) \delta_{vv'} \delta_{cc'} + i \Xi_{vc,v'c'}
  , \\
  H^C_{vc,c'v'} &=&  i \Xi_{vc,c'v'}
  ,
\end{eqnarray*}
where $v$ and $c$ run over, respectively, the occupied and empty $GW$ states $\varepsilon_i, \varphi_i(\vec{r})$.
Also, $\Xi = \delta\Sigma/\delta G$ is the Bethe-Salpeter kernel that, for singlet and triplet states, can be written as
\begin{eqnarray}
  ^1 \Xi_{ij,i'j'} &=& i \langle \varphi_j \varphi_{j'}^* | W | \varphi_i \varphi_{i'}^* \rangle - 2 i \langle \varphi_i^* \varphi_j | w | \varphi^*_{i'} \varphi_{j'} \rangle
  , \label{Xi1} \\
  ^3 \Xi_{ij,i'j'} &=&  i \langle \varphi_j \varphi_{j'}^* | W | \varphi_i \varphi_{i'}^* \rangle
  , \label{Xi2}
\end{eqnarray}
with $w(\vec{r},\vec{r}') = 1 / |\vec{r} - \vec{r'}|$ being the bare Coulomb interaction.
The excitonic eigenvalues $E_\lambda$ of the BSE equivalent equation~(\ref{bseeq}) are the final excitation energies of the system, while the eigenvectors $\Psi_\lambda$ are the excitonic wave functions from which one can extract the excitation oscillator strength.
In the Tamm-Dancoff approximation (TDA), the non-Hermitian coupling part is neglected, $ H^C = 0 $. 

In addition to $GW$ and BSE, we have also performed a time-dependent Hartree-Fock (TDHF) calculation, a common approximation in quantum chemistry, condensed matter, and frequently called RPA in the context of nuclear physics \cite{RingSchuck} (not to be confused with the RPA approximation in condensed matter).
TDHF is equivalent to a BSE calculation started from HF instead than $GW$, and taking $W=w$ (the bare Coulombian) in the BSE kernel Eqs.~(\ref{Xi1}) and (\ref{Xi2}).

We used a \emph{d-aug}-cc-pV5Z \cite{Dunning94} correlation-consistent Gaussian basis set with angular momentum up to $l=5$ and including a double set of diffuse orbitals.
HF calculations were carried by the \texttt{NWCHEM} package \cite{nwchem}, and $GW$ and BSE by the \texttt{Fiesta} code \cite{BlaseOlevano11,Jac15a,Li16} that integrates Eq.~(\ref{sigmagw}) by contour deformation and uses a Coulomb-fitting resolution of the identity (RI-V) with the associated auxiliary basis \emph{d-aug}-cc-pV5Z-RI \cite{Wei06}.

\begin{table}
 \begin{tabular}{lccrr}
   \hline
   $nl$ & HF & $GW$ & Exact \& EXP & Exact-DFT\\
   \hline
   $1s$ ($= - \mathrm{IP}$) & $-0.9143$ & $-0.9075$ & $-0.9037$ & $-0.9037$ \\
   $2s$ ($= - \mathrm{EA}$) & $+0.0217$ & $+0.0213$ & $> 0$ & $-0.1577$ \\
   $2p$ & $+0.0956$ & $+0.0944$ & & $-0.1266$ \\
   $3s$ & $+0.1394$ & $+0.1369$ & & $-0.0645$ \\
   \hline
 \end{tabular}
 \caption{He electron removal (first line) and addition (following lines) energies (Ha) in HF, $GW$, exact \cite{KonoHattori84} and experimental (EXP) result, and DFT-exact KS energies \cite{SavinGonze98}.}
 \label{qpenergies}
\end{table}

\paragraph{\textbf{Results - }}
In Table~\ref{qpenergies} we report our calculated He electron removal and addition HF and $GW$ quasiparticle (QP) energies, in comparison to the exact energies \cite{KonoHattori84} which coincide with the experiment within $10^{-4}$ Ha.
For the $1s$ removal energy, equal to minus the ionization potential (IP), the HF result presents an error of 0.3 eV which is reduced to only 0.1 eV in $GW$.
There are no exact calculations of the $2s$ addition energy, but it is known from the experiment that the He electron affinity (EA) is negative, so that the addition of a third electron costs energy and three-electron states are unbound.
Both HF and $GW$ correctly present a spectrum of unbound states (positive energy) in the electron addition part of the spectrum.
This is not the case of DFT Kohn-Sham (KS) energies, even the \textit{exact} ones.
Exact KS eigenvalues and the exact DFT exchange-correlation potential are available in He \cite{UmrigarGonze98,SavinGonze98} by inversion of the Kohn-Sham equation imposing the exact solution \cite{KonoHattori84}.
The last occupied exact KS eigenvalue is known \cite{PerdewBalduz82,LevySahni84,AlmbladhVonBarth85} to coincide with minus the ionization potential, and this is the case here for the $1s$.
However, the rest of the KS spectrum has no physical interpretation, and here we see that even \textit{exact} KS energies are qualitatively off.
On the other hand, exact KS energy differences were found \cite{UmrigarGonze98,SavinGonze98} surprisingly closer to the neutral (e.g., optical) excitation spectrum than KS energies are to charged (e.g., EA, photoemission) excitations.

\begin{table}[t]
\begin{tabular}{cccccccc}
  \hline
  $n^S\!L$ & HF & $G_0W_0$ & $GW$ & BSE$_\mathrm{TDA}$ & BSE & \textbf{Exact} & TDHF \\
  \hline
 $2^3\!S$ & \multirow{2}{*}{0.9396} & \multirow{2}{*}{0.9297} & \multirow{2}{*}{0.9289} & 0.7288 & 0.7271 & \textbf{0.7285} & 0.7237 \\
 $2^1\!S$ & & & & 0.7689 & 0.7676 & \textbf{0.7578} & 0.7759 \\
 $2^3\!P$ & \multirow{2}{*}{1.0136} & \multirow{2}{*}{1.0028} & \multirow{2}{*}{1.0020} & 0.7728 & 0.7724 & \textbf{0.7706} & 0.7806 \\
 $2^1\!P$ & & & & 0.7897 & 0.7894 & \textbf{0.7799} & 0.7997 \\
 $3^3\!S$ & \multirow{2}{*}{1.0574} & \multirow{2}{*}{1.0453} & \multirow{2}{*}{1.0444} & 0.8432 & 0.8427 & \textbf{0.8350} & 0.8499 \\
 $3^1\!S$ & & & & 0.8648 & 0.8637 & \textbf{0.8425} & 0.8732 \\
 \hline
\end{tabular}
\caption{
   He excitation energies in atomic units (Ha).
   The 0 of the energy is set to the He ground state $1^1\!S$. 
   }
\label{excitations}
\end{table}

In Fig.~\ref{helevels} and Table~\ref{excitations}, we report the first six helium excitation energies from HF and $GW$ energy differences, corrected by the BSE calculation and compared to the exact result \cite{KonoHattori84}, which also coincides with the experiment within $10^{-4}$ Ha, and to TDHF.
We see that neutral excitations are qualitatively and quantitatively captured by BSE, which reaches a good overall agreement with the exact result.
The very concepts of screening, exciton and BSE electron-hole propagators are questionable and pushed to their extreme limit of application in He, and yet BSE \textit{systematically improves} upon TDHF.
The error is reduced by a factor of 2 or better when passing from TDHF to BSE, i.e., introducing the screening.
Although one might have expected severe self-interaction and self-screening problems in He, the error is only 0.04 eV for the first $2^3\!S$ excitation and degrades for higher excitations, up to 0.6~eV for $3^1\!S$.
This degradation is merely a finite basis effect.
Indeed, in Table~\ref{convergence} we show the convergence of selected states with respect to the size of the Gaussian basis sets, and, in particular, with respect to the presence of diffuse (augmented) orbitals.
The ionization potential $\mathrm{IP} = - \epsilon_{1s}$ is converged already at the level of a standard cc-pVTZ Gaussian basis \cite{Dunning94} (not shown).
This is because the $1s$ state  is highly localized and requires few Gaussians to be represented accurately.
Higher states get more and more delocalized and, consequently, require larger (more diffuse)  basis sets.
The first $2^3\!S$ neutral excitation is converged at an augmented basis (\textit{aug}-cc-pV5Z), while the intermediate $2^1\!P$ requires a double augmentation level (\textit{d-aug}-cc-pV5Z).
The results for $3^1\!S$ are clearly less converged and accurate. 
Higher states are not converged even at this high level of double augmentation, and we do not report them.
Anyway, states towards the continuum of hydrogenic He$^+$(1$s$) plus a free electron are less interesting for the study of the many-body electron-electron interaction and would require better adapted bases, e.g., plane waves.
The same also holds for double excitations \cite{RomanielloOnida09,SangalliMarini11,ElliottMaitra11}, which are much more interesting, but, for He, lie deep in the continuum \cite{HicksComer75}.
Our Gaussian calculation indicates an absolute accuracy of $GW$ and BSE calculations of 0.1 eV for the physically relevant low-lying many-body excitation spectra, independent of finite basis or pseudopotential errors.

\begin{table}[t]
 \begin{tabular}{lrrrr}
  \hline
    & cc-pV5Z & \textit{\hphantom{d-}aug}-cc-pV5Z & \textit{d-aug}-cc-pV5Z & Exact\hphantom{6}\\
  \hline
  $\epsilon_{1s}$ & $-0.9066$ & $-0.9076$ & $-0.9075$ & $-0.9037$\hphantom{6} \\
  $E_{2^3\!S}$    & 0.8538 & 0.7284 & 0.7271 & 0.7285\hphantom{6} \\
  $E_{2^1\!P}$    & 1.3345 & 0.8684 & 0.7894 & 0.7799\hphantom{6} \\
  $E_{3^1\!S}$    & 2.6041 & 1.1952 & 0.8637 & 0.8425\hphantom{6} \\
  $f_{1^1\!S \to 2^1\!P}$    & 1.7607 & 0.7272 & 0.2763 & 0.27616 \\
  \hline
 \end{tabular}
 \caption{Basis set convercenge of the $\epsilon_{1s} = - \mathrm{IP}$, selected excitation energies $E_{n^S\!L}$, and the oscillator strength $f$ of $2^1\!P$.}
 \label{convergence}
\end{table}

In the following, we discuss the comparison between BSE and TDDFT results. 
Very accurate TDDFT results are reported in Ref.~\cite{PetersilkaBurke00} (see Fig.~1 and Tables~I and~II).
However, these results fundamentally rely on the use of unapproximated, exact-DFT Kohn-Sham eigenvalues.
Exact-DFT Kohn-Sham energy differences \cite{UmrigarGonze98,SavinGonze98,PetersilkaBurke00} are in He already within 0.5~eV of the exact excitation energies.
Therefore, the TDDFT kernel has the easier task of only splitting singlet and triplet states, which is already well done by the adiabatic LDA standard approximation.
Exact Kohn-Sham eigenvalues are exceptionally available only for He and another few systems where the exact Kohn-Sham potential is known by reverse engineering from the exact solution.
If both the DFT and TDDFT calculations are consistently done using approximations, e.g., LDA, as in the more standard procedure, the results drastically change.
\textit{None} of the excitations are bound in DFT+TDDFT LDA~\cite{WassermanBurke03}, and GGA does not improve.
Because of the missing long-range $1/r$ decay in the exchange-correlation potential, there is no Rydberg series in LDA or GGA atoms.
These standard approximations of DFT TDDFT are not even qualitatively validated on this workbench.

The physical mechanisms behind our HF-$GW$-BSE scheme are different.
From Fig.~\ref{helevels}, we see that HF as well as $GW$ energy differences, contrary to exact-DFT KS, are substantially different from the exact neutral excitation energies. 
This is a general feature: $GW$ quasiparticle energies physically represent electron removal or addition charged excitations, e.g., the IP and the EA or band plots measured in photoemission.
Quasiparticle energy differences are physically distinct from neutral optical excitations because they miss the electron-hole interaction.
The latter is introduced only at the level of the BSE.
From the difference between $GW$ and BSE levels in Fig.~\ref{helevels}, one can see the importance of the \textit{e-h} interaction.
In contrast with TDDFT, the BSE kernel manages not only to split singlet and triplet states, but, more importantly, to bring $GW$ levels, which are placed far in the continuum, down to the discrete bound spectrum region.
Our calculation has not made use of any additional helium specific knowledge from the exact solution, like the use of exact-DFT KS energies, and relies only on standard, commonly accepted approximations.
Specifically, it is derived on top of the $GW$ approximation, $\Xi = \delta \Sigma_{GW} / \delta G$, it neglects a variational term in $\delta W/\delta G$, and its $W$ in Eqs.~(\ref{Xi1}) and (\ref{Xi2}) is taken to be static, $W(\omega=0)$.
Therefore, the present severe workbench on the helium limiting case represents a validation of an already broadly applied recipe.

In Table~\ref{excitations} we also quote the results of BSE in the TDA approximation.
In He, TDA introduces an error not larger than other approximations in this methodology.
However, we already know several cases \cite{MaMolteni09,MaMolteni10,GruningGonze09} where TDA breaks down.
In Table~\ref{excitations} we also report intermediate $G_0W_0$ results.
Their values are higher than the eigenvalues of self-consistent $GW$ by only 8$\sim$9 $ \times 10^{-4}$ Ha, and this difference roughly propagates through BSE to the final excitation energies.
We confirm the conclusion of Ref.~\cite{BlaseOlevano11} that, in isolated systems, there is no appreciable improvement from going beyond $G_0W_0$ when HF is used as a starting point.

\begin{table}[t]
\begin{tabular}{cccc|ccc}
  \hline
   & BSE & \textbf{Exact} & TDHF & HF & Exact-DFT & \\
  \hline
  $f_{1^1\!S \to 2^1\!P}$ & 0.2763 & \textbf{0.27616} & 0.2916 & 0.2009 & 0.3243 & $\! \! f_{1s \to 2p}$\\
  \hline
\end{tabular}
\caption{
   He first excitation oscillator strengths.
   }
 \label{oscillatorstrength}
\end{table}

Finally, our work also provides the possibility of quantifying the quality of wave functions, independently from energies, through excitation oscillator strengths $f$ \cite{KonoHattori84}.
In BSE, $f_\lambda$ results from an expression implying a sum over all electron-hole electric-dipole matrix elements between QP wave functions $\langle \varphi_c | e^{-iqr} | \varphi_v \rangle$ times the exciton wave function $\Psi_\lambda^{vc}$ \textit{e-h} coefficients.
In the inset of Fig.~\ref{helevels}, we plot electron and hole distributions from the excitonic wave function $\Psi_{2^1\!P}(r_h,r_e)$ for the $2^1\!P$ state.
In Table~\ref{oscillatorstrength} we report the oscillator strength $f_{1^1\!S \to 2^1\!P}$ of the transition from the ground state $1^1\!S$ to the $2^1\!P$ excited state.
Transitions to other states studied here are all forbidden by selection rules obeyed in our electric-dipole approximation BSE code.
Starting from the HF value of 0.2009, we find for BSE 0.2763, in quantitative agreement with the exact value of 0.27616~\cite{KonoHattori84} and improving the TDHF of 0.2916.
The exact-DFT KS result is 0.3243 \cite{AppelBurke03}, and TDLDA is also around that value \cite{WassermanBurke03}.
These values confirm the accuracy of our HF-$GW$-BSE approach and, specifically, the quality of exciton wave functions $\Psi_\lambda$.

It is worth mentioning a last interesting finding of our BSE calculation.
One expects that an $nL$ excitation has, in general, a dominant contribution from the $1s \to nl, l=L$ transition, like the graph in Fig.~\ref{helevels} suggests, plus an admixture of higher energy transitions introduced by the BSE kernel $\Xi$.
We found that this is not the case for the triplet $n^3\!S$ excitations.
The $1s \to 2s$ transition contributes only for 42\% to the $2^3\!S$ excitation, while a 52\% dominant contribution comes from $1s \to 3s$.
The reverse holds for the highest energy $3^3\!S$ excitation, which has a 58\% contribution from $1s \to 2s$ and only 38\% from the $1s \to 3s$ transition.

\paragraph{\textbf{Conclusions - }}
We have benchmarked {\em ab initio} many-body $GW$ and BSE against exact results of the helium atom.
The standard approximations behind $GW$ and BSE accurately capture the physics of this two-body only limiting case.
Excitation energies are in good agreement with the exact spectra.
The surprising agreement on the $f_{1^1\!S \to 2^1\!P}$ oscillator strength indicates a good description of the wave functions.

\begin{acknowledgments}
\paragraph{\textbf{Acknowledgements - }} We thank Peter Schuck for seminal discussions and continuous support.
J.~L. and X.~B. acknowledge funding from the European Union Horizon 2020 Programme for Research and Innovation under Grant No. 646176.
\end{acknowledgments}

\bibliography{he}

\end{document}